\documentclass[journal,twoside,web]{ieeecolor2}
\usepackage{generic_arxiv}
\usepackage{cite}
\usepackage{amsmath,amssymb,amsfonts}
\usepackage{algorithmic}
\usepackage{graphicx}
\usepackage{textcomp}
\def\BibTeX{{\rm B\kern-.05em{\sc i\kern-.025em b}\kern-.08em
    T\kern-.1667em\lower.7ex\hbox{E}\kern-.125emX}}
\usepackage{multirow}
\usepackage{stfloats}
\usepackage[caption=false, font=footnotesize]{subfig}

\graphicspath{{Images/}{Images/algorithms_hsto_events/}{Images/hsto_healthy_pd/}{Images/combination/}{Images/gait_cycle/}{Images/correlation/}{Images/fmcw_algorithm/}{Images/fmcw_signal/}{Images/bland_altman/}{Images/group_error/}{Images/network_setup/}{Images/node_design/}{Images/rel_error_box/}{Images/snr_effect_vicon/}}
	
\begin{document}

\title{Radar Network for Gait Monitoring: Technology and Validation}
\author{Ignacio E. López-Delgado*, 
	Víctor Navarro-López,
	Francisco Grandas-Pérez,
	Juan~I.~Godino-Llorente,
	Jesús Grajal,
    \thanks{This work was supported by the projects \hbox{PID2020-113979RB-C21}, \hbox{TED2021-131688B-I00}, \hbox{PID2021-128469OB-I00}, founded by \hbox{MCIN/AEI/10.13039/501100011033} and \hbox{NextGenerationEU/PRTR}. The work of Ignacio E. López-Delgado was in part supported by an FPU Fellowship granted by the Spanish Ministry of Innovation, Science and Universities: \hbox{FPU20/06405}. \textit{(Corresponding author: Ignacio E. López-Delgado.)}}
    \thanks{This work involved human subjects in its research. Approval of all ethical
  	and experimental procedures and protocols was granted by the Ethics Committee of the Hospital General Universitario Gregorio Marañón with the code PD-RADAR-05/2024 in accordance with the Spanish Ethical Review Act.}
    \thanks{Ignacio E. López-Delgado, Juan I. Godino-Llorente and Jesús Grajal are with the Information Processing and Telecommunications Center, Universidad Politécnica de Madrid, 28040 Madrid, Spain (ie.lopez@upm.es; ignacio.godino@upm.es; jesus.grajal@upm.es).}
    \thanks{Víctor Navarro-López is with the Movement Analysis, Biomechanics, Ergonomics, and Motor Control Laboratory, Rey Juan Carlos University, 28922 Alcorcón, Spain (victor.navarro@urjc.es).}
    \thanks{Francisco Grandas-Pérez is with the Movement Disorders Unit, Hospital General Universitario Gregorio Marañón, 28007 Madrid, Spain (francisco.grandas@salud.madrid.org).}
}                        

\maketitle

\begin{abstract}
In recent years, radar-based devices have emerged as an alternative approach for gait monitoring. However, the radar configuration and the algorithms used to extract the gait parameters often differ between contributions, lacking a systematic evaluation of the most appropriate setup. Additionally, radar-based studies often exclude motorically impaired subjects, leaving it unclear whether the existing algorithms are applicable to such populations.

In this paper, a radar network is developed and validated by monitoring the gait of five healthy individuals and three patients with Parkinson's disease. Six configurations and four algorithms were compared using Vicon\textsuperscript{®} as ground-truth to determine the most appropriate solution for gait monitoring. The most accurate stride velocity and distance in the state of the art were obtained orienting two nodes towards the feet and one towards the torso. Moreover, we show that analyzing the feet velocity increases the reliability of the temporal parameters, especially with aged or motorically impaired subjects.

The contribution is significant for the implementation of radar networks in clinical and domestic environments, as it addresses critical aspects concerning the radar network configuration and algorithms.
\end{abstract}


\section{Introduction}

\IEEEPARstart{G}{ait} can be analyzed to anticipate the detection of some diseases, study their progression, observe the effectiveness of treatments and anticipate potential risks associated with aging~\cite{gurbuz_2024}. Nowadays, gait is usually evaluated following subjective assessments based on certain protocols, like the \textit{Timed Up-And-Go} (TUG)~\cite{podsiadlo_1991}. This subjective evaluation is typically carried out in clinical environments by specialists in movement disorders. 

Due to the limited capacity of health systems and the required expertise, these evaluations are generally reserved for patients with clinical signs of movement disorders and are not available for the general population. Additionally, these evaluations bias the gait of the patients~\cite{liu_2022}: the gait is modified by sensors attached to the body and by the pressure caused by clinical environments. 

In order to provide more objective evaluations, clinicians often use different techniques to collect and analyze gait. The most common ones are based on wearable sensors~\cite{delrobaei_2018, perumal_2016} and/or video cameras~\cite{zanela_2022}:

\begin{itemize}
	\item Wearable sensors, such as inertial measurement units (IMUs), are low-cost devices that provide accurate measurements with a small computational cost. However, they need to be attached to the body, so they can cause discomfort. Moreover, they cannot provide continuous measurements because their battery is limited~\cite{delrobaei_2018, perumal_2016}.
	\item Video cameras also provide very accurate measurements. However, they invade the subject's privacy, are vulnerable to obstacles and low-illumination environments, have a high computational cost, and require expensive devices~\cite{zanela_2022}. The current gold-standard in gait monitoring integrates infra-red cameras with wearable reflectors. This is the case of the Vicon\textsuperscript{®} system~\cite{vicon}.
\end{itemize}

Despite the objectiveness provided by these techniques, their limitations significantly hinder their implementation in home or clinical settings. On the contrary, radar devices open the door to a potential monitoring of gait in a pervasive, continuous and non-obstructive way~\cite{liu_2022,siva_2024,seifert_2019}.

Among other advantages, radar devices are low-cost, contactless and can provide continuous measurements preserving the subject's privacy without causing discomfort~\cite{liu_2022}. The main contributions to the state-of-the-art are summarized in Table~\ref{tab:state_of_the_art}~\cite{liu_2022, kabelac_2019, saho_2022, soubra_2023, gurbuz_2024, wang_2014, siva_2024, abedi_2022, hadjipanayi_2024, zeng_2022, wang_2024, lopezdelgado_2025}.

\begin{table*}[htbp]
	\caption{Comparison of Radar Systems Implemented for Gait Monitoring without Using Treadmills}
	\begin{center}
		\begin{tabular}{lccc}
			\hline 
			\textbf{Contribution}				& \textbf{Architecture}				& \textbf{Node focus}& \textbf{Implementation}						\\
			\hline
			Wang et al. \cite{wang_2014}		& Multi-node						& 1 torso, 1 feet		& Clinical and domestic hallways			\\
			Kabelac et al. \cite{kabelac_2019}	& Single-node, MIMO					& 1 torso				& Clinical and domestic environments		\\
			Liu et al. \cite{liu_2022}			& Single-node, MIMO 				& 1 torso				& Clinical and domestic environments		\\
			Saho et al. \cite{saho_2022}		& Single-node						& 1 torso, 1 feet		& Clinical and domestic hallways			\\
			Abedi et al. \cite{abedi_2022}		& Single-node, MIMO 				& 1 torso				& Clinical and domestic hallways			\\
			Zeng et al. \cite{zeng_2022}		& Single-node, MIMO 				& 1 feet				& Clinical and domestic hallways			\\
			Soubra et al. \cite{soubra_2023}	& Single-node						& 1 torso + feet		& Clinical and domestic hallways			\\
			Gurbuz et al. \cite{gurbuz_2024}	& Single-node, MIMO 				& 1 torso				& Clinical and domestic hallways			\\
			Siva et al. \cite{siva_2024}		& Multi-node, MIMO 					& 3 torso				& Clinical and domestic environments		\\
			Hadjipanayi et al. \cite{hadjipanayi_2024}		& Multi-node, MIMO 		& 3 torso + feet		& Clinical and domestic hallways			\\
			Wang et al. \cite{wang_2024}		& Single-node, MIMO 				& 1 torso, 1 feet		& Clinical and domestic hallways			\\
			López-Delgado et al. \cite{lopezdelgado_2025}		& Single-node, MIMO & 1 feet				& Clinical and domestic hallways			\\
			This work							& Distributed multi-node			& 2 torso, 2 feet		& Clinical and domestic hallways 			\\ 
			\hline    
		\end{tabular}
		\label{tab:state_of_the_art}
	\end{center}
	\vspace{-0.5cm}
\end{table*}

Several radars have been validated with infra-red cameras, accelerometers and ground pressure sensors~\cite{liu_2022, saho_2022, soubra_2023, siva_2024, gurbuz_2024, hadjipanayi_2024, zeng_2022, wang_2024}. However, direct comparisons between these contributions are challenging, as they use different radar configurations (number and location of radars) and algorithms, not addressing which are optimal~\cite{gurbuz_2024}. 

Regarding the radar network configuration, some works have implemented single-node systems~\cite{liu_2022, kabelac_2019, saho_2022, soubra_2023, gurbuz_2024, zeng_2022, wang_2024}, which are usually limited to constrained gait paths. Fewer works have focused on implementing multi-node systems, achieving higher accuracies and opening the door to gait monitoring in unconstrained situations~\cite{siva_2024, hadjipanayi_2024}.

The optimal radar location is also being discussed. One option is to focus only at the torso motion using torso radars~\cite{liu_2022, kabelac_2019, saho_2022, gurbuz_2024, siva_2024, abedi_2022, wang_2024}. The algorithms presented in~\cite{saho_2022} extrapolate the heel-strike (HS) and toe-off (TO) events from the torso velocity. Alternatively, torso radars can extract the gait velocity by tracking a person along the room, following the algorithms presented in~\cite{liu_2022}.

Another option is to focus on the feet motion with feet radars featuring the algorithms proposed in~\cite{seifert_2019}. Feet radars are effective at extracting temporal parameters, but they encounter difficulties in extracting the torso velocity and can be occluded at homes~\cite{zeng_2022, seifert_2019, lopezdelgado_2025}.

In order to exploit the advantages of each different configuration, they can be deployed simultaneously~\cite{saho_2022, soubra_2023, wang_2014, hadjipanayi_2024}. However, the question of whether the increase of the system complexity is a worthwhile pursuit, given the occasional increment in system accuracy, remains a topic of debate~\cite{saho_2022}.

Furthermore, motorically impaired subjects are often left out of radar validation studies, not addressing whether the algorithms and configurations implemented are applicable for such populations. For instance, torso radars can be inefficient for aged or motorically impaired subjects because the velocity of the torso is affected by certain movement disorders~\cite{trabassi_2023}. Moreover, they face challenges in reliably estimating the foot maximum velocity, critical to assess gait asymmetries related with several diseases, like Parkinson's Disease (PD)~\cite{giannakou_2023}.

In this paper, we present an approach to improve radar-based gait analysis by validating a configurable radar network using a Vicon system as ground-truth~\cite{vicon}. The validation is carried out analyzing large gait trials (more than ten minutes) of five healthy individuals and three patients with PD. To the best of our knowledge, this is the first radar network validation that considers different health conditions\footnote{Although the work presented in~\cite{siva_2024} was deployed at homes as a radar network, the validation was carried out with a single node.}.

Six different radar network configurations (alternating single-node, multi-node, torso and feet deployments) and three different algorithms are implemented to extract a set of gait parameters~\cite{liu_2022, seifert_2019, saho_2022}. The performance of the different implementations is compared to choose the most suited option.

The results show for the first time that feet radars are more accurate at extracting temporal gait parameters when analyzing aged or motorically impaired subjects. Furthermore, the most accurate stride velocities of the state-of-the-art are extracted with the radar network configuration presented, demonstrating its suitability for clinical or domestic implementations.

This paper is structured as follows: Section~\ref{sec:radar_node} presents a radar node designed for gait monitoring. The radar network is validated in Section~\ref{sec:radar_network}, analyzing the gait of eight individuals presenting different health conditions. Section~\ref{sec:discussion} analyzes the overall performance of this radar network, addressing key considerations for in-home implementations. Finally, Section \ref{sec:conclusion} presents the final conclusions of this work.

\section{24-GHz Radar Node}\label{sec:radar_node}

The radar node shown in Fig.~\ref{fig:node_info} is designed for home and clinical deployments. It features the following characteristics:

\begin{figure}[htbp]
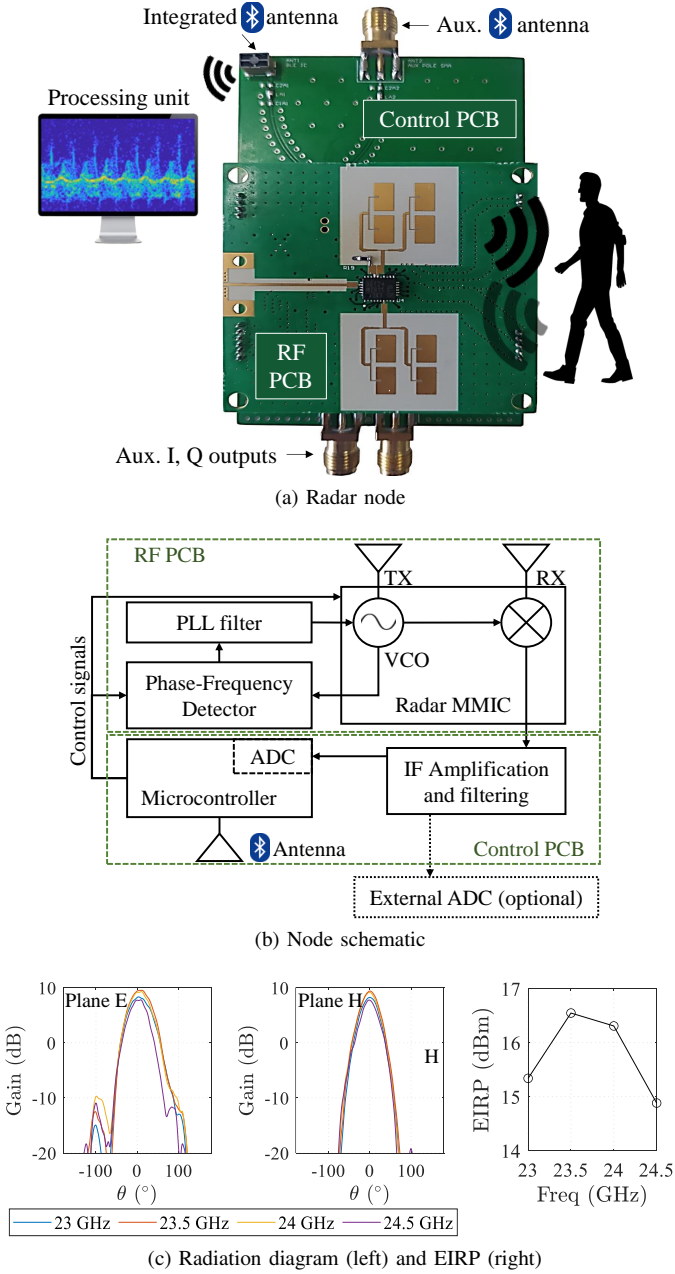

	\centering
	\subfloat[Radar node]{\includegraphics[width=0.9\linewidth]{node_pcb}
	\label{fig:node_pcb}}
	\hfil
	\subfloat[Node schematic]{\includegraphics[width=0.8\linewidth]{node_schematic}
	\label{fig:node_schematic}}
	\hfil
	\subfloat[Radiation diagram (left) and EIRP (right)]{\includegraphics[width=\linewidth]{antenna_diagram_2d}
	\label{fig:antenna_diagram_2d}}
	\hfil
	\hspace{0.02\textwidth}
	\caption{24-GHz radar node. (a) RF and Control PCBs with the external elements that interact with the radar. (b) Schematic of the node. (c) Antenna characterization (left) and radar EIRP (right).}
	\label{fig:node_info}
\end{figure}

\begin{itemize}
	\item Compact design: 60x80 mm$^2$.
	\item Full gait analysis: The antennas feature a 40$^\circ$ beamwidth in both planes (as shown in Fig.~\ref{fig:antenna_diagram_2d}), to monitor individuals walking 1 m away from the radar. The node operative parameters are configurable (see Table~\ref{tab:node_requirements}~\cite{lopez_2023}).
	\item Wireless communication protocol: The nodes can send the recorded data to a processing unit using Bluetooth\textsuperscript{®} Low Energy~\cite{stm32wb15cc}.
	\item Low-cost system: The nodes use commercial-off-the-shelf components and microstrip antennas. The total cost of the prototype shown in Fig.~\ref{fig:node_schematic} is $\sim$ 150 €.
	\item Scalable design: The nodes are easily integrated into a radar network, as shown in Section \ref{sec:radar_network}.
\end{itemize}

\begin{table}[t]
	\caption{Radar Node Operative Parameters}
	\begin{center}
		\begin{tabular}{lc}
			\hline 
			\textbf{Parameter} 								& \textbf{Operative range}		\\
			\hline
			Operational modes								& CW-Doppler, LFMCW				\\
			Range (m)										& 1 - 8								\\
			Beamwidth, $\vartheta$ ($^\circ$)	    		& 40								\\
			Central frequency, $f_0$ (GHz) 					& 22.5 - 25.5						\\
			Transmitted power, $P_{TX}$ (dBm)				& (-2) - (7) 							\\
			Antenna gain, $G$ (dB)							& 9									\\
			Bandwidth, $B$ (GHz)								& 0 - 2							\\
			Chirp time, $T_c$ (ms)							& 0.1 - 18							\\
			Sampling rate, $f_s$ (ksps)					 	& 125 - 500							\\
			Continuous interrogation						& Yes
			\\ \hline    
		\end{tabular}
		\label{tab:node_requirements}
	\end{center}
\end{table}

The radar nodes have two boards connected: the RF PCB and the control PCB, as shown in Fig.~\ref{fig:node_schematic}. 

The RF PCB contains the radar MMIC Infineon\textsuperscript{®} BGT24MTR11~\cite{bgt24mtr11}, the phase-frequency detector (PFD) Analog Devices\textsuperscript{®} ADF4159~\cite{adf4159}, and the antennas. The characterization of the antennas, carried out in an anaechoic chamber, is shown in Fig.~\ref{fig:antenna_diagram_2d}. The maximum EIRP considering the antenna gain and the MMIC output power is 15.5 $\pm$ 1 dBm.

The Control PCB contains the power control unit, the IF signal conditioning stage, and the microcontroller STM32WB15CC~\cite{stm32wb15cc} that samples the IQ signals and sends them via Bluetooth. Alternatively, the IQ signals can be sampled with an external Analog to Digital converter (ADC), as is done in this work. 

\begin{figure}[t]
    \centering
	\includegraphics[width=\linewidth]{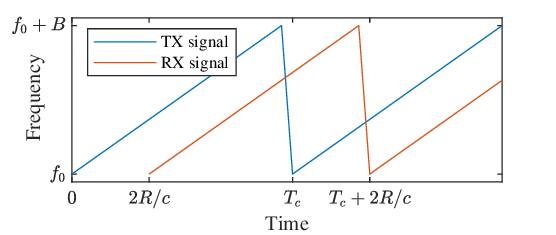}			
	\caption{Signal model of LFMCW radars, where $f_0$ is the start frequency, $B$, the bandwidth, $T_c$, the chirp time, $R$, the range of the target and $c$, the speed of light.}
	\label{fig:fmcw_signal}
\end{figure}

The radar node is configured in Linear Frequency-Modulated Continuous-Wave (LFMCW) mode. The signal model of the LFMCW radar is shown in Fig.~\ref{fig:fmcw_signal}, where it is shown that the frequency difference between the transmitted and received signals (beat frequency) is proportional to the range $R$ of the target. The IQ signals result from low-pass filtering the mix of the transmitted and received signals.

\begin{figure}[t]
    \centering
	\includegraphics[width=0.85\linewidth]{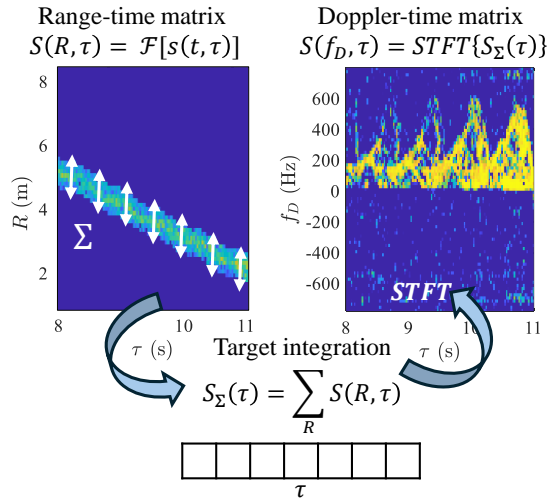}
	\caption{Algorithm implemented to extract the range-time and Doppler-time matrices.}
	\label{fig:algorithm_fmcw}
\end{figure}

The range-time and Doppler-time matrices are used to extract the gait parameters~\cite{adib_2015, liu_2022, tivive_2015, seifert_2019, ash_2018, hu_2022}, as will be explained in Sections~\ref{sec:hs_to_events} to~\ref{sec:other_parameters}. The steps followed to extract these matrices (see Fig.~\ref{fig:algorithm_fmcw}) are:

\begin{enumerate}
	\item Extract the range-time matrix $S \left(R, \tau\right)$, calculating the Fast-Fourier Transform with a Hann window of the IQ signals: $S \left(R, \tau\right) = \mathcal{F}\left[s \left(t, \tau\right)\right]$. The range-time matrix is filtered in the time dimension to eliminate the clutter using a high-pass filter with 10-Hz cutoff frequency~\cite{adib_2015, ash_2018}. The transient state is eliminated from the measurement.
	\item Integrate the bins of the target: The range bins containing the target, represented with white arrows in Fig~\ref{fig:algorithm_fmcw}, are added coherently: $S_{\Sigma} \left(\tau\right) = \sum_{R} S\left(R, \tau\right)$~\cite{adib_2015, hu_2022}.
	\item Extract the Doppler-time matrix: The Doppler-time matrix $S \left(f_D, \tau\right)$ is calculated extracting the short-time Fourier Transform (STFT) of $S_{\Sigma} \left(\tau\right)$: $S \left(f_D, \tau\right) = \mathrm{STFT}\left\{S_{\Sigma} \left(\tau\right)\right\}$, using a $T_M = $ 50 ms Hann window with a time shift of one sample.
\end{enumerate}

\section{Radar Network Validation}\label{sec:radar_network}

\begin{figure}[t]
	\centering
	\subfloat[]{\includegraphics[width=\linewidth]{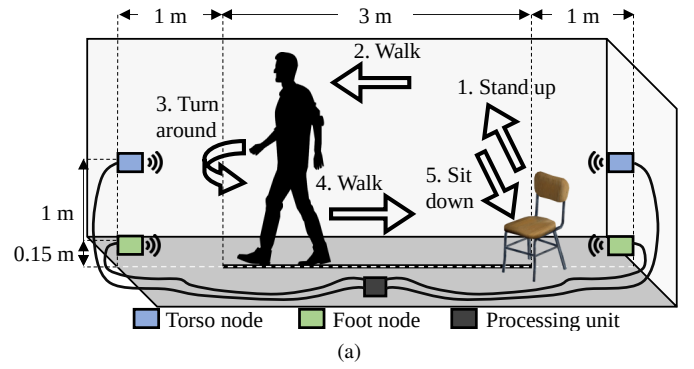}
		\label{fig:network_architecture}}
	\hfil
	\subfloat[]{\includegraphics[width=0.8\linewidth]{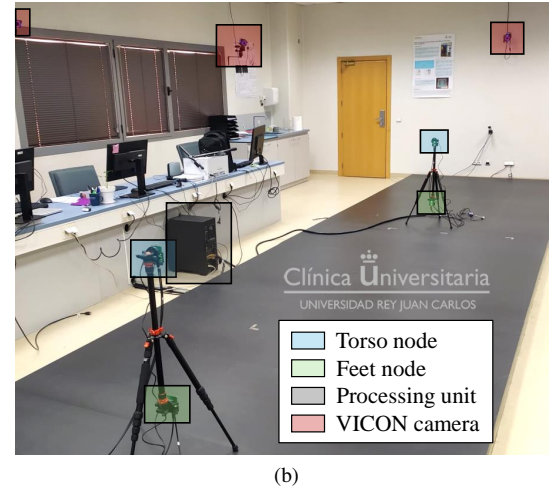}
		\label{fig:setup_lambecom}}
	\caption{Radar network implementation. (a) Illustration of the radar network with a subject performing the TUG test. (b) Photograph of the experimental setup highlighting the nodes and the Vicon cameras used as ground-truth.}
	\label{fig:network_implementation}
\end{figure}

The radar network implemented in this study has been configured for analyzing subjects undergoing clinical gait assessments. Two tripods, each holding a feet node and a torso node, are placed facing one another at the ends of the walking platform, as shown in Fig.~\ref{fig:network_implementation}. The four nodes connected to an external processing unit.

Comparisons are carried out using a Vicon system equipped with 8-Valkyrie cameras as ground-truth, synchronized with the radar's ADC using a trigger signal generated by the Vicon device. The participants hold 32 Vicon markers placed on the sternum, 10$^\mathrm{th}$ thoracic vertebra, left and right interior and exterior wrists, anterior and posterior superior iliac spines, thighs, knees, tibiae, ankles, heels and toes. 

\begin{figure}[t]
	\centering
	\includegraphics[width=0.8\linewidth]{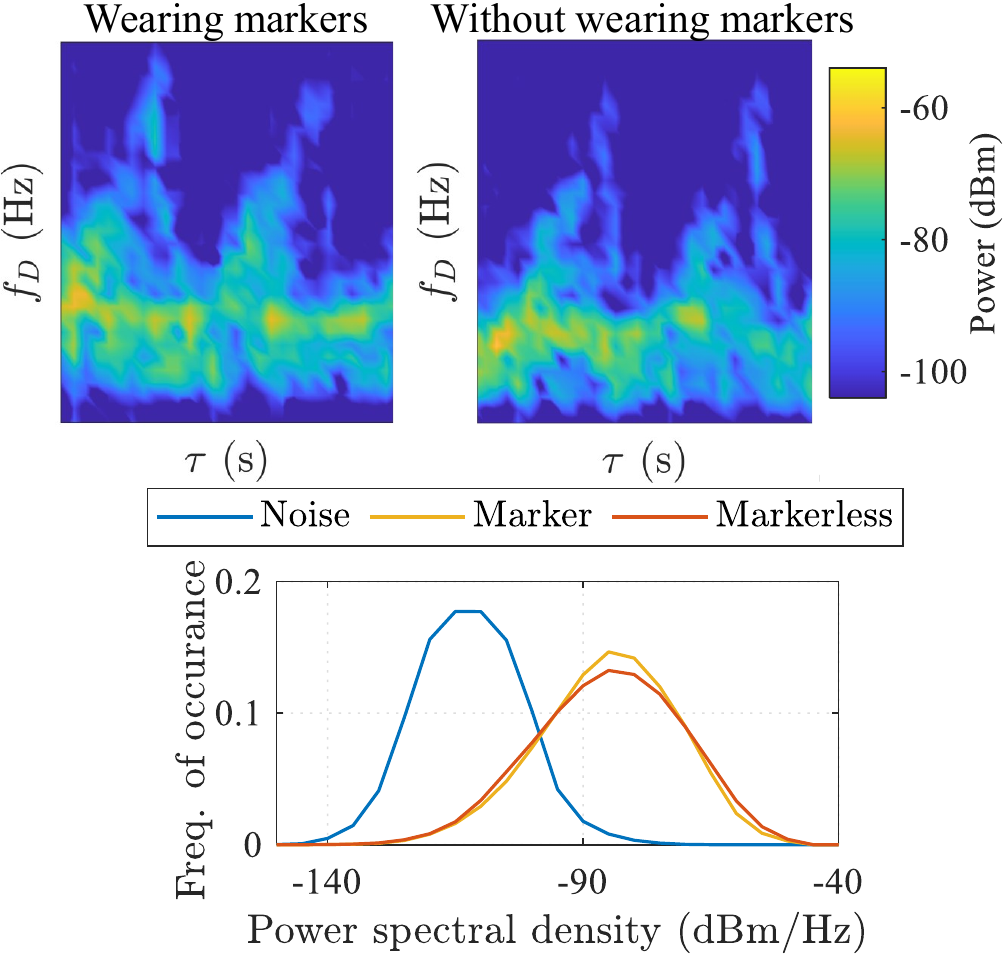}			
	\caption{Vicon wearable markers have minimal impact on the SNR of the Doppler-time matrix. The signal power distribution with and without them is shown in the bottom figure, displaying a similar SNR in both cases.}
	\label{fig:snr_effect_vicon}
\end{figure}

Experiments were carried out to verify that the Vicon markers do not enhance reflections at 24 GHz. To achieve this, a person with and without wearing them is analyzed. The results shown in Fig.~\ref{fig:snr_effect_vicon} suggest that their impact is minimal. Thus, the validation presented in Sections~\ref{sec:hs_to_events}~to~\ref{sec:other_parameters} can be extended to scenarios where the person wears no markers.

The signals captured by the Vicon are processed as explained in~\cite{oconnor_2007}, as in other radar validations~\cite{hadjipanayi_2024}. Nonetheless, the techniques presented in~\cite{oconnor_2007} report temporal errors of $\pm$25~ms~\cite{oconnor_2007}, as they focus on the feet velocity rather than on the feet location.

The four radar nodes are configured with $f_0$~=~23~GHz, $B$~=~1.4~GHz and $T_c$~=~625~$\mu$s~\cite{lopez_2023}. They capture signals simultaneously, but can be processed independently. This way, we compare the radar configurations shown in Table~\ref{tab:configuration_nodes}: C1-C4 are compared in this section. The results obtained are then discussed in Section~\ref{sec:discussion} to analyze C5 and C6.

\begin{table}[t]
	\centering
	\caption{Configurations Compared to Evaluate the Performance of the Radar Network and the Signal Processing Algorithms}
	\label{tab:configuration_nodes}
	\begin{tabular}{@{}ccc@{}}
		\hline
		Configuration & Torso nodes & Feet nodes \\ \hline
		Configuration 1 (C1)           & 1           & 0          \\
		Configuration 2 (C2)           & 0           & 1          \\
		Configuration 3 (C3)           & 2           & 0          \\
		Configuration 4 (C4)           & 0           & 2          \\
        Configuration 5 (C5)           & 1           & 2          \\
        Configuration 6 (C6)           & 2           & 2          \\ \hline
		\end{tabular}
\end{table}

\begin{figure*}[t]
	\centering
	\subfloat[]{\includegraphics[width=0.3\linewidth]{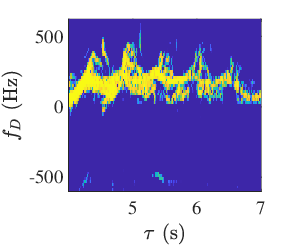}
	\label{fig:torso_1}}
	\hfil
	\subfloat[]{\includegraphics[width=0.3\linewidth]{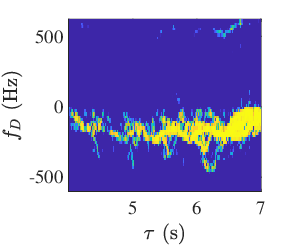}
	\label{fig:torso_2}}
	\hfil
	\subfloat[]{\includegraphics[width=0.3\linewidth]{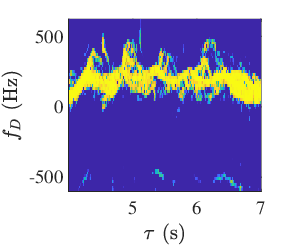}
	\label{fig:torso_comb}}
	\hfil
	\caption{Combination of the Doppler-time matrices extracted with two torso nodes. (a) Node 1, (b) Node 2, (c) Combination of nodes 1 and 2. The combined Doppler-time matrix displays the time frames of the node with the highest SNR: node 1 for $\tau < 5.5 \,\mathrm{s}$, and node 2 for $\tau > 5.5 \,\mathrm{s}$. Prior to the combination, the Doppler-time matrix of node 2 is flipped with respect to the Doppler axis.}
	\label{fig:comb_udoppler}
\end{figure*}

C1 and C2 experience an SNR decay when the subject walks away from the radar. To handle this, the Doppler signatures of the nodes with the same height in C3 and C4 are combined by selecting the time frames of the Doppler-time matrix of the node with higher SNR, as shown in Fig.~\ref{fig:comb_udoppler}. Prior to this combination, one of the Doppler-time matrices is flipped with respect to the Doppler axis. As a result of this combination, it is obtained a single Doppler-time matrix with enhanced SNR. 

Four signal processing algorithms are compared for each configuration. Two algorithms estimate the temporal gait parameters (stride, step, stance, swing and double-support times): one analyzing the feet motion~\cite{seifert_2019}, and the other the torso velocity~\cite{saho_2022}. The remaining two algorithms extract the spatial parameters (stride and step distance and velocity): one from the range-time matrix~\cite{liu_2022}, and the other from the torso velocity~\cite{seifert_2019}.

\subsection{Protocol and Corpus}

The radar network along with the Vicon system were deployed at the Laboratorio de Biomecánica, Ergonomía y Control Motor (LAMBECOM) belonging to the Universidad Rey Juan Carlos I, Madrid, Spain.

Participants were recruited by the Neurology Service of Hospital General Universitario Gregorio Marañón (HGUGM), Madrid, Spain. All participants were properly informed about the experiment conditions and agreed to participate by signing a consent form. Participants did not receive any compensation, and were informed of their rights and the possibility of leaving the study at any time. Patients were individually identified with a code, which is different from the one used for their clinical histories. No personal data was exchanged with the researchers who processed and curated the corpus.

Eight participants are recorded performing clinical trials:

\begin{itemize}
	\item TUG~\cite{podsiadlo_1991} (trial 1): The subject stands up from a chair, walks 3~m in a straight line, turns around, walks back to the chair and sits down, as shown in Fig.~\ref{fig:network_architecture}. The participants perform the TUG ten times at quick pace (compared to each participant's normal pace). 
	\item Continuous walk: The subject walks in a straight line and turns around every 3~m. The participants are asked to perform four different continuous-walk trials: Twenty repetitions at normal pace (trial~2), slow pace (trial~3), quick pace (trial~4), and 120~s at normal pace (trial~5).
\end{itemize}

Although the entire trials are recorded, the velocity data extracted from the torso Vicon markers is processed to separate the gait sequences~\cite{soubra_2023} and discard the turning, standing and sitting stages.

The aforementioned trials guarantee that the radar network is validated at different gait speeds, and that it is robust against gait changes like standing, turning and sitting.

The participants are divided into three groups according to their age and health condition:

\begin{itemize}
	\item Group 1 (G1): Healthy and young participants; 2 male and 1 female, aged between 21 and 24 years old without known motor disorders.
	\item Group 2 (G2): Healthy and aged participants; 1 male and 1 female, aged between 55 and 71 years old without known motor disorders.
	\item Group 3 (G3): PD participants; 2 male and 1 female, aged between 59 and 78 years old with Parkinson's Disease diagnosed in the last four years previous to the experiment.
\end{itemize}

Although the number of participants is scarce, it is important to consider that each participant is recorded walking for more than ten minutes. Thus, a large number of gait cycles per participant is analyzed, larger than other state-of-the-art contributions \cite{saho_2022, soubra_2023, siva_2024, gurbuz_2024, hadjipanayi_2024}, as shown in next sections. 

\subsection{Validation: Heel-Strike and Toe-Off Events}\label{sec:hs_to_events}

\begin{figure}[t]
	\centering
	\includegraphics[width=\linewidth]{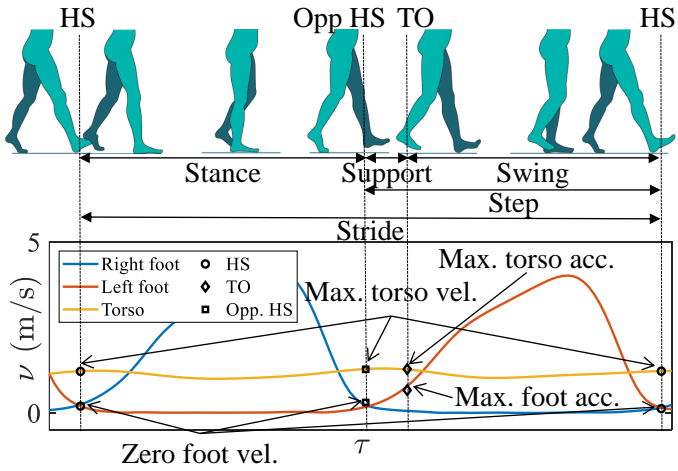}			
	\caption{Gait cycle stages and events extracted from a Vicon motion capture. The HS and TO events are reflected in the velocity of the feet and torso.}
	\label{fig:gait_cycle}
\end{figure}

A gait cycle, represented in Fig.~\ref{fig:gait_cycle}, is the interval between two heel strikes (HS) of the primary foot. After the first HS, the opposite heel lays on the ground (opposite HS event), and the primary foot leaves the ground (toe-off (TO) event).

\begin{figure}[t]
	\centering
	\subfloat[]{\includegraphics[width=0.48\linewidth]{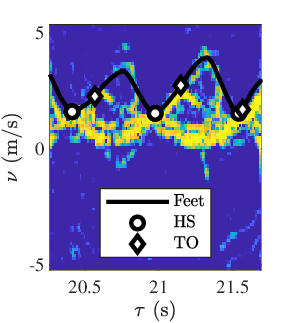}
	\label{fig:alg_seifert}}
	\hfil
	\subfloat[]{\includegraphics[width=0.48\linewidth]{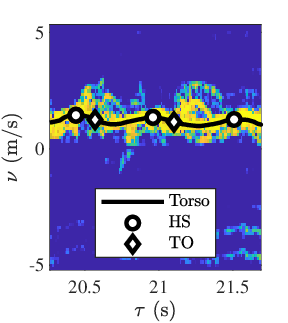}
	\label{fig:alg_saho}}
	\caption{Illustration of the algorithms implemented to identify the HS and TO events from the Doppler contributions of the feet (a) and the torso (b).}
	\label{fig:algorithm_comparison}
\end{figure}

There are two approaches to determine the HS and TO events: 
\begin{itemize}
	\item The method proposed by Seifert et al.~\cite{seifert_2019} is based on the fact that the velocity of the feet is zero when they lay on the ground, as shown in Fig.~\ref{fig:gait_cycle}. The HS events are the local minima of the feet Doppler frequency. The Doppler contribution of the feet is isolated because they present the largest velocity, as shown in Fig.~\ref{fig:alg_seifert}. This method is considered more efficient for feet nodes, so it is applied for C2 and C4.
	\item The method proposed by Saho et al.~\cite{saho_2022} is based on the correlation between the torso velocity and the HS and TO events, shown in Fig.~\ref{fig:gait_cycle}. This method identifies the HS events at the torso Doppler frequency peaks, and the TO events at the torso acceleration peaks. The Doppler frequency of the torso is easily isolated because the torso represents the most powerful Doppler component, as shown in Fig.~\ref{fig:alg_saho}. This method is considered more efficient for torso nodes, so it is applied for C1 and C3.
\end{itemize}

The HS detection ratio, shown in Table~\ref{tab:hs_detection_ratio}, measures the ratio of HS events detected by both the radar and the Vicon, and the HS events detected by the Vicon (ground-truth). As a reference, Table~\ref{tab:hs_detection_ratio} also includes the number of gait cycles. It is important to consider that the first and last gait cycles of each sequence are discarded and not considered in Table~\ref{tab:hs_detection_ratio}. Four main conclusions are drawn out from Table~\ref{tab:hs_detection_ratio}:

\begin{enumerate}
    \item C4, featuring 2 feet nodes, presents the highest HS detection ratio for all the tests carried out.
    \item Adding a second feet node (comparing C2 with C4) increases the HS detection ratio, while adding a second torso node (comparing C1 with C3) does not. This is because the SNR decay experienced when the subjects move away from the radar is stronger in the case of the feet, because the torso is the largest reflector.
    \item Feet nodes are robust for all gait velocities. On the contrary, the performance of torso nodes decays when the subject walks fast (tests 1 and 4).
\end{enumerate}

\begin{table}[t]
	\centering
	\caption{HS Detection Ratio. The Last Column Represents Number of Gait Cycles}
	\label{tab:hs_detection_ratio}
	\begin{tabular}{@{}lccccc@{}}
		\hline
		    & C1   & C2   & C3   & C4   & Gait cycles \\ \hline
		Test 1 & 0.81 & 0.95 & 0.78 & \textbf{0.98} & 192 \\
		Test 2 & 0.87 & 0.97 & 0.85 & \textbf{0.99} & 412 \\
		Test 3 & 0.94 & 0.96 & 0.94 & \textbf{0.99} & 505 \\
		Test 4 & 0.68 & 0.94 & 0.75 & \textbf{0.99} & 320 \\
		Test 5 & 0.88 & 0.95 & 0.89 & \textbf{0.99} & 525 \\ \hline
		All    & 0.86 & 0.96 & 0.86 & \textbf{0.99} & 1954 \\ \hline
	\end{tabular}
\end{table}


The effect of the radar configuration is also analyzed for aged and motorically impaired subjects. 
A strong correlation between the torso velocity and the HS and TO events is reported for healthy subjects. However, this relationship is lost for motorically impaired subjects, as a consequence of a movement disorder. Therefore, it is more convenient to extract the HS and TO events by looking at the feet motion, contrary to the proposal of~\cite{saho_2022}.

\begin{figure}[t]
\centering
\subfloat{\includegraphics[width=0.48\textwidth]{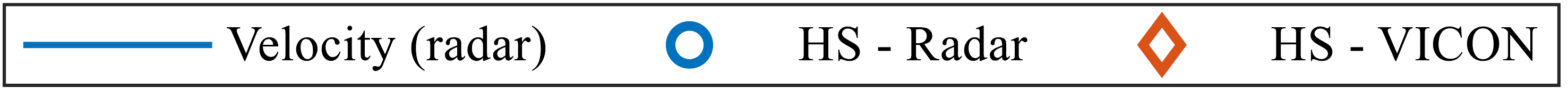}}
\hfil
\setcounter{subfigure}{0}
\subfloat[]{\includegraphics[width=0.23\textwidth]{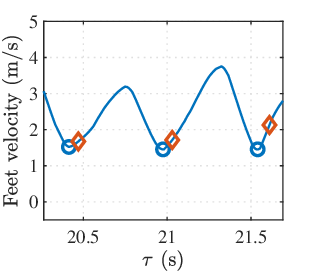}
\label{fig:temporal_feet_healthy}}
\hfil
\subfloat[]{\includegraphics[width=0.23\textwidth]{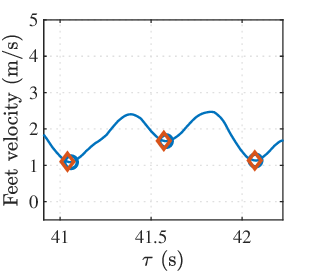}
\label{fig:temporal_feet_disease}}
\hfil
\subfloat[]{\includegraphics[width=0.23\textwidth]{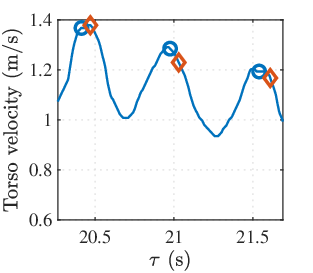}
\label{fig:temporal_torso_healthy}}
\hfil
\subfloat[]{\includegraphics[width=0.23\textwidth]{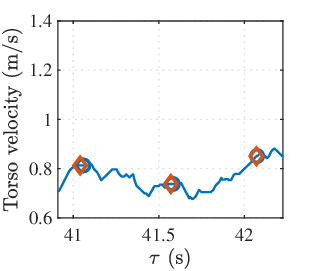}
\label{fig:temporal_torso_disease}}
\caption{Feet motion (a and b) is reliable to extract the HS and TO events for both healthy and motorically impaired subjects. The torso motion (c and d) is not reliable for motorically impaired subjects (d). All figures show a gait cycle, comparing the HS events extracted with the Vicon (ground-truth) and the radar.}
\label{fig:feet_algorithms_comp}
\end{figure}

This effect is observed in Fig.~\ref{fig:feet_algorithms_comp}: Figs. \ref{fig:feet_algorithms_comp}a and b show the feet velocity of two people of G1 and G3, respectively, performing Test 5. In both cases, the HS events are clearly differentiable, i.e. the local minima of the feet velocity~\cite{seifert_2019}. Figs. \ref{fig:feet_algorithms_comp}c and d show the torso velocity during the same time frames, exhibiting a pattern that makes the detection of the HS events (i.e., the local maxima of the torso velocity~\cite{saho_2022}) more difficult for those participants belonging to G3.

\subsection{Temporal Gait Parameters}\label{sec:temp_parameters}

The HS and TO events are used to compute five gait parameters, represented in Fig.~\ref{fig:gait_cycle}: 
\begin{itemize}
	\item Stride time: between HS events of the same foot.
	\item Step time: between HS events of different feet.
	\item Stance time: between HS and TO events of the same foot.
	\item Swing time: between TO and HS events of the same foot.
	\item Double-support time: between HS and TO events of different feet.
\end{itemize}

For each gait parameter extracted by both the radar ($P_{\mathrm{rad}}$) and the Vicon ($P_{\mathrm{vic}}$), the absolute error $|\varepsilon|$ and the absolute value of the relative error $|\varepsilon_r|$ are extracted, given by Eqs.~\ref{eq:abs_err} and~\ref{eq:abs_rel_err}, respectively. These metrics were used in~\cite{hadjipanayi_2024, wang_2024}.

\begin{equation}\label{eq:abs_err}
	|\varepsilon| = \left|P_{\mathrm{vic}}-P_{\mathrm{rad}} \right|
\end{equation}

\begin{equation}\label{eq:abs_rel_err}
	|\varepsilon_r| = \left| \cfrac{P_{\mathrm{vic}}-P_{\mathrm{rad}}}{P_{\mathrm{vic}}} \right|
\end{equation}

\begin{figure*}[t]
\centering
\subfloat{\includegraphics[width=0.19\textwidth]{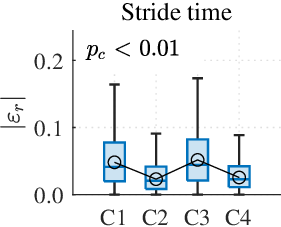}
\label{fig:box_rel_t_stride}}
\hfil
\subfloat{\includegraphics[width=0.19\textwidth]{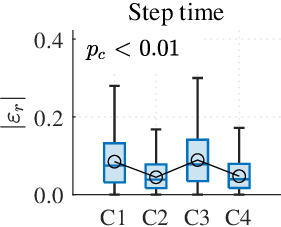}
\label{fig:box_rel_t_step}}
\hfil
\subfloat{\includegraphics[width=0.19\textwidth]{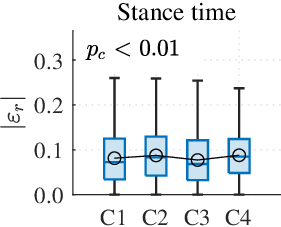}
\label{fig:box_rel_t_stance}}
\hfil
\subfloat{\includegraphics[width=0.19\textwidth]{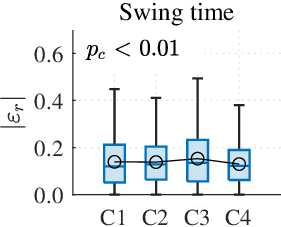}
\label{fig:box_rel_t_swing}}
\hfil
\subfloat{\includegraphics[width=0.19\textwidth]{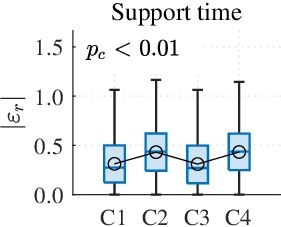}
\label{fig:box_rel_t_support}}
\caption{Distribution of $|\varepsilon_r|$ for the different radar network configurations considering the temporal gait parameters. The mean of $|\varepsilon_r|$ is represented with a black circle. When $p_c < 0.01$, the configuration with the smallest error is significantly better.}
\label{fig:rel_error_temp_box}
\end{figure*}

Radar network configurations are compared aiming to:

\begin{itemize}
	\item Maximize the HS detection ratio (see Section~\ref{sec:hs_to_events}).
	\item Minimize the error of the temporal parameters: Fig.~\ref{fig:rel_error_temp_box} shows the distribution of $|\varepsilon_r|$ (considering all subject groups). The Kruskal-Wallis $p$-value across configurations, $p_c$, shows statistical differences between them~\cite{kruskall_1952}: if $p_c < 0.01$, the configuration with the smallest error is significantly better.
	\item Ensure independence from the subject's health, which should not affect the accuracy. This dependency is evaluated through the Kruskal-Wallis $p$-value across subject groups, $p_g$~\cite{kruskall_1952}, shown in Table~\ref{tab:mean_error_temp} together with the mean of $|\varepsilon|$ and $|\varepsilon_r|$: configurations with $p_g > 0.01$ are unaffected by the health of the subject.
\end{itemize}

Fig.~\ref{fig:rel_error_temp_box} and Table~\ref{tab:mean_error_temp} show that:

\begin{itemize}
	\item Two feet-node radar networks (C4) provide significantly more reliable temporal metrics than two torso nodes (C3). This is because feet radars extract the HS and TO events directly from the feet velocity. In fact, the accuracy of C4 is withing the order of magnitude of the ground-truth's accuracy: 25~ms~\cite{oconnor_2007}.
	\item C4 is equally accurate for all subject groups considering the stride, step and stance time, but presents differences for the swing and double-support times.
	\item Smaller temporal parameters (swing and double-support times) present larger $|\varepsilon_r|$. The reason is that $|\varepsilon|$ is similar for all the temporal parameters, i.e. a 0.05~s error in a 1.2~s stride leads to $|\varepsilon_r|$=0.04, while the same 0.05~s error in a 0.44 s swing phase leads to $|\varepsilon_r|$=0.13.
\end{itemize}

\noindent These conclusions are further discussed in Section \ref{sec:discussion}.


\begin{table}[t]
	\centering
	\caption{Temporal Parameters Figures of Merit}
	\label{tab:mean_error_temp}
	\begin{tabular}{lcccccc}
		\hline
		Conf.               & Metric & Stride$^\dagger$        & Step$^\dagger$         & Stance$^\dagger$       & Swing$^\dagger$        & Support$^\dagger$      \\ \hline
		\multirow{3}{*}{C1}
& $|\varepsilon_r|$ & 0.06 & 0.10 & 0.09 & 0.15 & \textbf{0.37} \\ 
& $|\varepsilon|$ & 65 & 55 & 62 & 64 & 41 \\ 
& $p_g$ & $>$ 0.01 & $>$ 0.01 & $<$ 0.01 & $>$ 0.01 & \textbf{$>$ 0.01} \\ 
\hline 
\multirow{3}{*}{C2}
& $|\varepsilon_r|$ & 0.04 & 0.06 & 0.09 & 0.15 & 0.45 \\ 
& $|\varepsilon|$ & 40 & 35 & 62 & 63 & 54 \\ 
& $p_g$ & $>$ 0.01 & $<$ 0.01 & $<$ 0.01 & $<$ 0.01 & $>$ 0.01 \\ 
\hline 
\multirow{3}{*}{C3}
& $|\varepsilon_r|$ & 0.06 & 0.10 & 0.09 & 0.16 & 0.38 \\ 
& $|\varepsilon|$ & 68 & 58 & 59 & 69 & 41 \\ 
& $p_g$ & $>$ 0.01 & $>$ 0.01 & $<$ 0.01 & $>$ 0.01 & $>$ 0.01 \\ 
\hline  
\multirow{3}{*}{C4}
& $|\varepsilon_r|$ & \textbf{0.03} & \textbf{0.06} & \textbf{0.09} & \textbf{0.14} & 0.45 \\ 
& $|\varepsilon|$ & 37 & 34 & 61 & 61 & 53 \\ 
& $p_g$ & \textbf{$>$ 0.01} & \textbf{$>$ 0.01} & \textbf{$>$ 0.01} & $<$ 0.01 & $<$ 0.01 \\ 
\hline 

		& \multicolumn{2}{l}{$^\dagger$$|\varepsilon|$ in ms} & \multicolumn{4}{l}{$p_g$: Kruskal-Wallis $p$-value across groups} \\
	\end{tabular}
\end{table}

\subsection{Spatial Gait Parameters}

The spatial gait parameters are the stride distance and step distance, which measure the distance traveled during a stride or a step, respectively; and the stride velocity and step velocity, which measure the mean gait velocity during a stride or a step, respectively. There are two approaches to extract them:

\begin{itemize}
	\item The method proposed by Seifert et al.~\cite{seifert_2019} is based on calculating the mean torso velocity during a step or a stride to later calculate the distance multiplying by the temporal parameters. This method is implemented for C1 and C3, as is more efficient for torso nodes.
	\item Liu et al.~\cite{liu_2022} calculate the distance traveled from the range-time matrix, to later calculate the velocity dividing the distance by the temporal parameters. This method can be applied for both feet and torso nodes, being the only adequate method for feet nodes. Thus, it is implemented for C2 and C4.
\end{itemize}

\begin{figure}[t]
\centering
\subfloat{\includegraphics[width=0.19\textwidth]{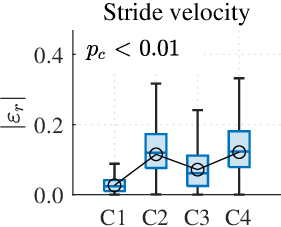}
\label{fig:box_rel_v_stride}}
\hfil
\subfloat{\includegraphics[width=0.19\textwidth]{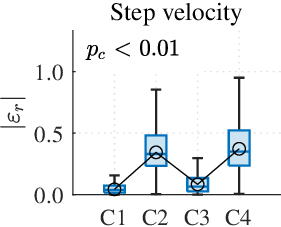}
\label{fig:box_rel_v_step}}
\hfil
\subfloat{\includegraphics[width=0.19\textwidth]{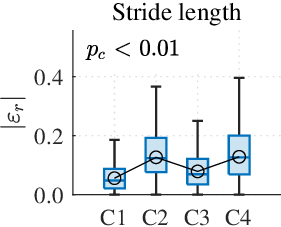}
\label{fig:box_rel_l_stride_torso}}
\hfil
\subfloat{\includegraphics[width=0.19\textwidth]{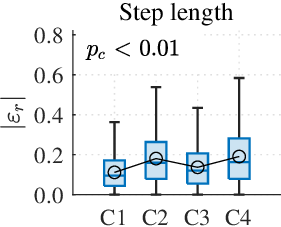}
\label{fig:box_rel_l_step_torso}}
\caption{Distribution of $|\varepsilon_r|$ for the different radar network configurations considering the spatial gait parameters. The mean of $|\varepsilon_r|$ is represented with a black circle. When $p_c < 0.01$, the configuration with the smallest error is significantly better.}
\label{fig:rel_error_dist_box}
\end{figure}

As in Section~\ref{sec:temp_parameters}, radar network configurations are compared to minimize the error and ensure independence from the subject's health. Fig.~\ref{fig:rel_error_dist_box} shows the distribution of $|\varepsilon_r|$ for the spatial gait parameters and the Kruskal-Wallis $p$-value across configurations, $p_c$~\cite{kruskall_1952}. Moreover, Table~\ref{tab:mean_error_dist} shows the mean of $|\varepsilon|$ and $|\varepsilon_r|$, and the Kruskal-Wallis $p$-value across subject groups, $p_g$.

Fig.~\ref{fig:rel_error_dist_box} and Table~\ref{tab:mean_error_dist} show that the torso radars (C1 and C3) are more accurate extracting spatial parameters. In fact, C1 is the best configuration, as deploying two torso nodes (C3) slightly degrades the performance of the system. This may be caused because the Doppler-time matrix combination introduces errors in the Doppler component of the torso. This conclusion is further discussed in Section~\ref{sec:discussion}.

\begin{table}[t]
	\centering
	\caption{Spatial Parameters Figures of Merit.}
	\label{tab:mean_error_dist}
	\begin{tabular}{lccccc}
		\hline
		Conf.               & Metric & Str. vel.\footnotesize$^\ddagger$     & Step vel.\footnotesize$^\ddagger$     & Str. dist.\footnotesize$^*$    & Step dist.\footnotesize$^*$    \\ \hline
		\multirow{3}{*}{C1}
& $|\varepsilon_r|$ & \textbf{0.03} & \textbf{0.06} & \textbf{0.06} & \textbf{0.12} \\ 
& $|\varepsilon|$ & 28 & 51 & 65 & 60 \\ 
& $p_g$ & \textbf{$>$ 0.01} & $<$ 0.01 & \textbf{$>$ 0.01} & $<$ 0.01 \\ 
\hline 
\multirow{3}{*}{C2}
& $|\varepsilon_r|$ & 0.15 & 0.40 & 0.16 & 0.19 \\ 
& $|\varepsilon|$ & 139 & 348 & 160 & 92 \\ 
& $p_g$ & $<$ 0.01 & $>$ 0.01 & $<$ 0.01 & $<$ 0.01 \\ 
\hline 
\multirow{3}{*}{C3}
& $|\varepsilon_r|$ & 0.07 & 0.09 & 0.09 & 0.14 \\ 
& $|\varepsilon|$ & 67 & 80 & 90 & 71 \\ 
& $p_g$ & $<$ 0.01 & $<$ 0.01 & $<$ 0.01 & $<$ 0.01 \\ 
\hline 
\multirow{3}{*}{C4}
& $|\varepsilon_r|$ & 0.15 & 0.42 & 0.16 & 0.20 \\ 
& $|\varepsilon|$ & 141 & 356 & 156 & 94 \\ 
& $p_g$ & $<$ 0.01 & $>$ 0.01 & $<$ 0.01 & $<$ 0.01 \\ 
\hline 

		& \multicolumn{2}{l}{$^\ddagger$$|\varepsilon|$ in mm/s} & \multicolumn{2}{l}{$^*$$|\varepsilon|$ in mm} & \\
		& \multicolumn{4}{l}{$p_g$: Kruskal-Wallis $p$-value across groups}& \\
	\end{tabular}
\end{table}

\subsection{Other Spatiotemporal Gait Parameters}\label{sec:other_parameters}

There are other spatiotemporal gait parameters that can be extracted with some configurations proposed in this study.

The cadence is a commonly-used parameter that measures the number of steps per minute \cite{giannakou_2023}. This parameter is not analyzed in this study as can be extracted as the inverse of the step time, which has already been reported.

The foot maximum velocity is a gait parameter critical for assessing gait asymmetries related with some diseases, such as PD~\cite{giannakou_2023, lopezdelgado_2025}. The best results are obtained with C4: $|\varepsilon_r|$~=~0.10, $|\varepsilon|$~=~34~cm/s, $p_g < 0.01$.

\section{Discussion}\label{sec:discussion}

The previous section showed that a three-node radar network featuring one torso node and two feet nodes is the most adequate between the ones studied to extract all the spatiotemporal gait parameters in a clinical or a hallway implementation. This configuration, named as C5, is even more accurate than a four-node configuration, as shown in Table~\ref{tab:mean_error_opt}. This section addresses some considerations concerning these configurations.

\subsection{Radar Network Reliability} 

\begin{figure*}[t]
\centering
\subfloat{\includegraphics[width=0.19\textwidth]{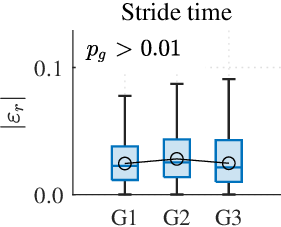}
	\label{fig:group_t_stride}}
\hfil
\subfloat{\includegraphics[width=0.19\textwidth]{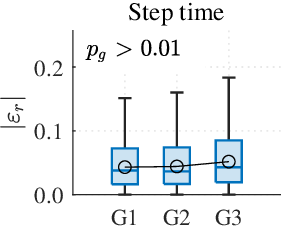}
	\label{fig:group_t_step}}
\hfil
\subfloat{\includegraphics[width=0.19\textwidth]{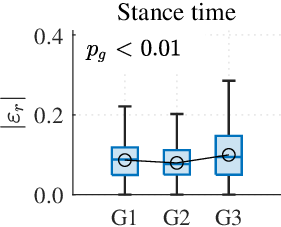}
	\label{fig:group_t_stance}}
\hfil
\subfloat{\includegraphics[width=0.19\textwidth]{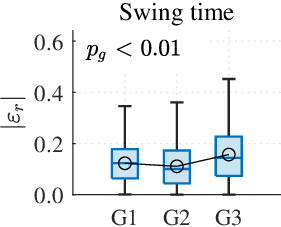}
	\label{fig:group_t_swing}}
\hfil
\subfloat{\includegraphics[width=0.19\textwidth]{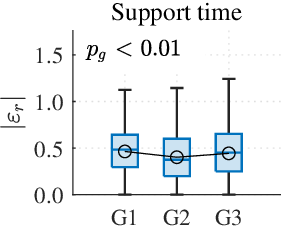}
	\label{fig:group_t_support}}
\hfil
\subfloat{\includegraphics[width=0.19\textwidth]{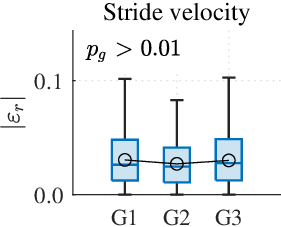}
	\label{fig:group_v_stride}}
\hfil
\subfloat{\includegraphics[width=0.19\textwidth]{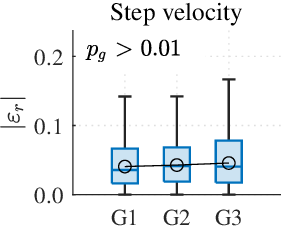}
	\label{fig:group_v_step}}
\hfil
\subfloat{\includegraphics[width=0.19\textwidth]{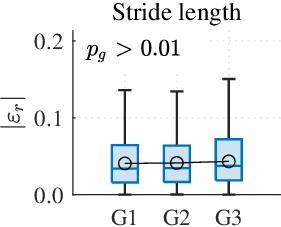}
	\label{fig:group_l_stride_torso}}
\hfil
\subfloat{\includegraphics[width=0.19\textwidth]{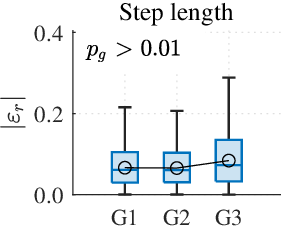}
	\label{fig:group_l_step_torso}}
\hfil
\subfloat{\includegraphics[width=0.19\textwidth]{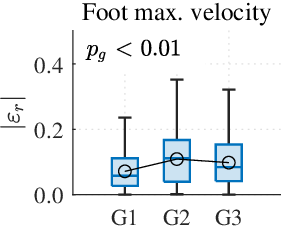}
	\label{fig:group_v_max_foot}}
\caption{Comparison of the relative absolute error of the different subject groups using two feet nodes and one torso node (C5). The mean accuracy of each parameter with each group is represented in black. The radar network is unaffected by the subject's health when $p_g > 0.01$.}
\label{fig:group_error}
\end{figure*}

\begin{table*}[t]
	\centering
	\caption{Figures of Merit for C5 and C6 (One Torso and Two Feet Nodes, and Two Torso and Two Feet Nodes, Respectively)}
	\label{tab:mean_error_opt}
	\begin{tabular}{@{}lccccccccccc@{}}
		\hline
	Conf.       & Metric & Stride t.$^\dagger$     & Step t.$^\dagger$       & Stance t.$^\dagger$     & Swing t.$^\dagger$      & Support t.$^\dagger$    & Stride vel.$^\ddagger$   & Step vel.$^\ddagger$     & Stride dist.$^*$  & Step dist.$^*$    & Feet vel.$^\mathsection$ \\ \hline
	\multirow{2}{*}{C5}
& $|\varepsilon_r|$ & \textbf{0.03} & \textbf{0.06} & \textbf{0.09} & \textbf{0.14} & \textbf{0.48} & \textbf{0.04} & \textbf{0.06} & \textbf{0.05} & \textbf{0.09} & \textbf{0.10} \\ 
& $|\varepsilon|$ & 36 & 33 & 63 & 63 & 55 & 35 & 50 & 53 & 44 & 35 \\ 
\hline 
\multirow{2}{*}{C6}
& $|\varepsilon_r|$ & 0.03 & 0.06 & 0.09 & 0.14 & 0.48 & 0.08 & 0.10 & 0.09 & 0.12 & 0.10 \\ 
& $|\varepsilon|$ & 36 & 34 & 63 & 63 & 55 & 72 & 82 & 85 & 56 & 35 \\ 
\hline 

	& \multicolumn{2}{l}{$^\dagger$$|\varepsilon|$ in ms} & \multicolumn{2}{l}{$^\ddagger$$|\varepsilon|$ in mm/s} & \multicolumn{2}{l}{$^*$$|\varepsilon|$ in mm} & \multicolumn{2}{l}{$^\mathsection$$|\varepsilon|$ in cm/s} & & & \\
    \end{tabular}
\end{table*}
\begin{table*}[t]
	\centering
	\caption{Validation of Radar Systems for Gait Analysis. A Gait Segment Involves Walking for few Meters. The Error Metrics Are: Mean Error \cite{saho_2022}, Mean-Squared Error \cite{soubra_2023}, Mean Absolute Error \cite{hadjipanayi_2024, zeng_2022}, and Standard Deviation \cite{siva_2024, gurbuz_2024, wang_2024}. Most Accurate Results are Represented in Bold$^{\mathrm{2}}$}
	\label{tab:comp_validation}
	\begin{tabular}{lcccc}
		\hline
		\textbf{Contribution}            & \textbf{Ground-truth}                                             & \textbf{Participants}                                                    & \textbf{Gait segments} & \textbf{Gait parameters (error)}                                                                                                                                                                                                             \\ \hline
		Saho et al. \cite{saho_2022}     & Vicon                                                             & \begin{tabular}[c]{@{}c@{}}10, age: 21-23, \\ healthy\end{tabular}            & 100                    & \begin{tabular}[c]{@{}c@{}}Stride t. (1\%), step t. (1\%), step dist. (2\%),\\ stance t. (18\%), swing t. (11\%) \end{tabular}                                                                                                       \\ \hline
		Zeng et al. \cite{zeng_2022}  & \begin{tabular}[c]{@{}c@{}}Accelerometer\\ (Xsens\textsuperscript{®})\end{tabular}                                              & \begin{tabular}[c]{@{}c@{}}3, age: 20-40, \\ healthy \end{tabular} & 60                    & Step t. (4\%) \\ \hline 
		Soubra et al. \cite{soubra_2023} & Vicon                                                             & \begin{tabular}[c]{@{}c@{}}26, age: 22-60,\\ healthy\end{tabular}        & 468                    & \begin{tabular}[c]{@{}c@{}}Stride t. (4\%), step t. (4\%), \textbf{swing t. (6\%)}, \\ gait vel. (3\%), stride dist. (4\%), step dist. (3\%)\end{tabular}                                                                \\ \hline
		Siva et al. \cite{siva_2024}     & \begin{tabular}[c]{@{}c@{}}Pressure sensors\\ (Zeno\textsuperscript{®})\end{tabular} & \begin{tabular}[c]{@{}c@{}}35, age: 60-89,\\ \textbf{frail}\end{tabular}          & 625                    & Step dist. (10\%)                                                                                                                                                                                                                           \\ \hline
		Gurbuz et al. \cite{gurbuz_2024} & Vicon                                                             & \begin{tabular}[c]{@{}c@{}}5, age: 23-27,\\ healthy\end{tabular}              & $\sim$300                   & \textbf{Stride t. (2\%)}, stride dist. (5\%)                                                                                                                                                                                                       \\ \hline
		Hadjipanayi et al. \cite{hadjipanayi_2024}  & Vicon                                              & \begin{tabular}[c]{@{}c@{}}12, age: 21-27, \\ healthy \end{tabular} & 72                    & \begin{tabular}[c]{@{}c@{}}Stride t. ($<$5\%), \textbf{step t. ($<$5\%)}, swing t. ($<$9\%), \\ \textbf{stance t. ($<$6\%), double-support t. ($<$43\%),} \\ stride vel. ($<$5\%), \textbf{stride dist. ($<$5\%), step dist. ($<$10\%)}, \end{tabular} \\ \hline
		 Wang et al. \cite{wang_2024}  & \begin{tabular}[c]{@{}c@{}}Infra-red cameras\\ (Qualisys\textsuperscript{®})  \end{tabular}                                            & \begin{tabular}[c]{@{}c@{}}9, age: 23-26, \\ healthy \end{tabular} & 360                    & \begin{tabular}[c]{@{}c@{}}Stride t. (4\%), step t. (7\%), stride vel. (5\%), \\  \textbf{step vel. (5\%)}, \textbf{foot max. vel. (5\%)}  \end{tabular} \\ \hline 
		This work                        & Vicon                                                             & \begin{tabular}[c]{@{}c@{}}8, age: 21-78, \\ healthy and \textbf{PD}\end{tabular} & \textbf{780}                    & \begin{tabular}[c]{@{}c@{}}Stride t. (3\%), step t. (6\%), stance t. (10\%), swing t. (15\%), \\ double-support t. (50\%), \textbf{stride vel. (4\%)}, step vel. (6\%), \\ \textbf{stride dist. (5\%)}, \textbf{step dist. (9\%)}, foot max. vel. (11\%) \end{tabular} \\ \hline
	\end{tabular}
\end{table*}

Table~\ref{tab:mean_error_opt} shows the error distribution of~C5, and a four-node radar network, named as~C6. Besides for the double-support times, the mean $|\varepsilon_r|$ is below 15\% for all gait parameters reported. Furthermore, C5's accuracy is mostly independent of the subject's health condition, as reflected by the Kruskal-Wallis $p$-value across subject groups, $p_g$, shown in Fig.~\ref{fig:group_error}.
	
Table~\ref{tab:comp_validation} compares the errors reported with C5 and other state-of-the-art contributions\footnote{The errors extracted in this work are not compared to the errors in~\cite{saho_2022} and the distance errors in~\cite{soubra_2023}. The former computed the mean error after averaging different trials, while the latter assumed the total distance walked is known, being the results hereby presented independent of the total distance.}:

\begin{itemize}
    \item This contribution reports the most accurate stride velocity, stride distance, and step distance metrics. Moreover, it is one of the few to report the accuracy of the foot maximum velocity, which can be used to analyze gait symmetry ~\cite{wang_2024, giannakou_2023, lopezdelgado_2025}.
	\item The most accurate stride and swing time results are reported by Gurbuz~et~al.~\cite{gurbuz_2024} and Soubra~et~al.~\cite{soubra_2023}, respectively, while Hadjipanayi~et~al.~\cite{hadjipanayi_2024} reported the most accurate step, stance and double-support times. Wang~et~al.~\cite{wang_2024} reported the most accurate step and maximum foot velocities. These validations were not carried out with motorically impaired subjects. Thus, in~\cite{gurbuz_2024, soubra_2023}, it is possible to foresee an accuracy decay when analyzing such subjects, as only torso radars were used, and we showed that their temporal accuracy decays for the subjects belonging to G3 (see Fig.~\ref{fig:feet_algorithms_comp}). As regards~\cite{hadjipanayi_2024,wang_2024}, the accuracy achieved is in agreement with the methods proposed in this paper, as they focus on the movement of the feet, and are close to the ground-truth accuracy.
    \item Although some state-of-the-art validations are carried out with more participants, the validation presented in this work evaluates more gait cycles, considering also motorically impaired subjects. Still, it would be convenient to confirm the results obtained in this work with more measurements.
\end{itemize}

\subsection{Considerations for Domestic Implementations}

\begin{figure*}[t]
\centering
\subfloat{\includegraphics[width=0.19\textwidth]{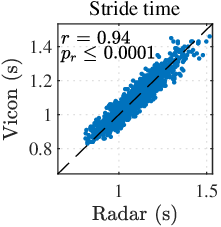}
	\label{fig:corr_t_stride}}
\subfloat{\includegraphics[width=0.19\textwidth]{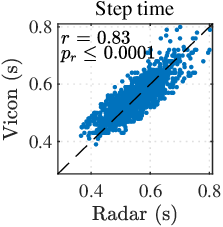}
	\label{fig:corr_t_step}}
\subfloat{\includegraphics[width=0.19\textwidth]{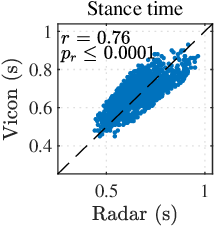}
	\label{fig:corr_t_stance}}
\subfloat{\includegraphics[width=0.19\textwidth]{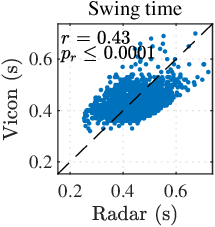}
	\label{fig:corr_t_swing}}
\subfloat{\includegraphics[width=0.21\textwidth, height=0.195\textwidth]{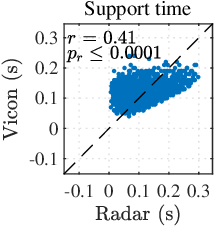}
	\label{fig:corr_t_support}}
\hfil
\subfloat{\includegraphics[width=0.19\textwidth]{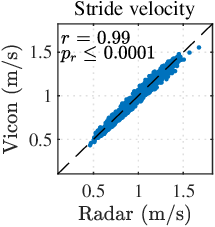}
	\label{fig:corr_v_stride}}
\hfil
\subfloat{\includegraphics[width=0.19\textwidth]{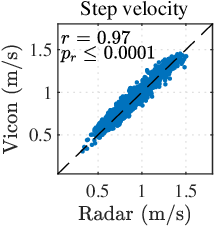}
	\label{fig:corr_v_step}}
\hfil
\subfloat{\includegraphics[width=0.19\textwidth]{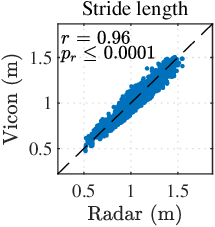}
	\label{fig:corr_l_stride_torso}}
\hfil
\subfloat{\includegraphics[width=0.19\textwidth]{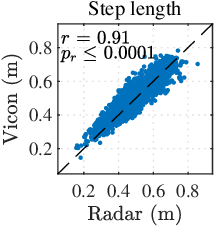}
	\label{fig:corr_l_step_torso}}
\hfil
\subfloat{\includegraphics[width=0.19\textwidth]{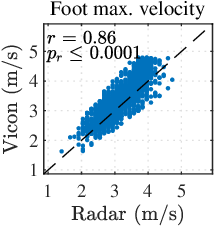}
	\label{fig:corr_v_max_foot}}
\caption{Correlation plots of the spatio-temporal gait parameters extracted with C5, showing correlation and the p-value coefficients, $r$ and $p_r$ respectively, and the ideal correlation case (dashed line).}
\label{fig:correlation}
\end{figure*}

\begin{figure*}[t]
\centering
\subfloat{\includegraphics[width=0.19\textwidth]{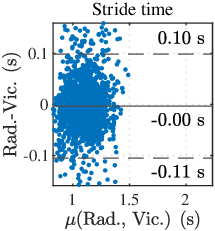}
	\label{fig:ba_t_stride}}
\hfil
\subfloat{\includegraphics[width=0.19\textwidth]{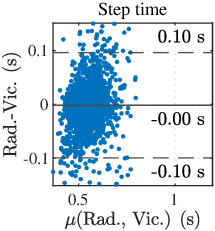}
	\label{fig:ba_t_step}}
\hfil
\subfloat{\includegraphics[width=0.19\textwidth]{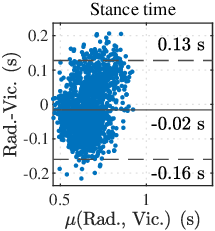}
	\label{fig:ba_t_stance}}
\hfil
\subfloat{\includegraphics[width=0.19\textwidth]{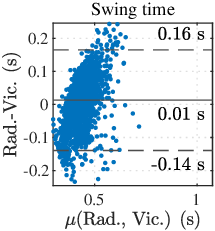}
	\label{fig:ba_t_swing}}
\hfil
\subfloat{\includegraphics[width=0.19\textwidth]{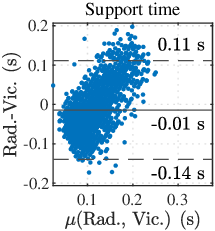}
	\label{fig:ba_t_support}}
\hfil
\subfloat{\includegraphics[width=0.19\textwidth]{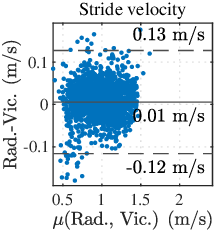}
	\label{fig:ba_v_stride}}
\hfil
\subfloat{\includegraphics[width=0.19\textwidth]{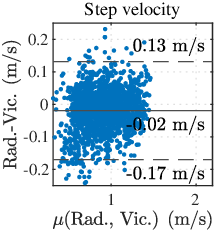}
	\label{fig:ba_v_step}}
\hfil
\subfloat{\includegraphics[width=0.19\textwidth]{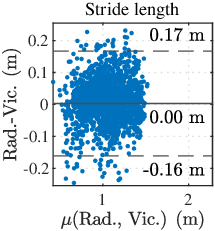}
	\label{fig:ba_l_stride_torso}}
\hfil
\subfloat{\includegraphics[width=0.19\textwidth]{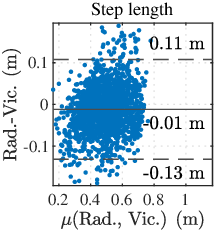}
	\label{fig:ba_l_step_torso}}
\hfil
\subfloat{\includegraphics[width=0.19\textwidth]{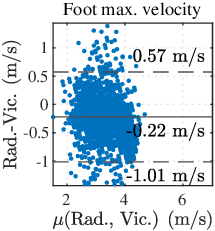}
	\label{fig:ba_v_max_foot}}
\caption{Bland-Altman plots of the spatio-temporal gait parameters extracted with C5, where $\mu\left(\mathrm{Rad.,\,Vic.}\right)$ represents the mean of each parameter extracted with the radar and the Vicon. The solid horizontal line represents the mean difference. The dashed horizontal lines represent the limits of agreement.}
\label{fig:bland_altman}
\end{figure*}

The radar network presented in this paper is suitable for analyzing gait in clinical and domestic scenarios. This claim is supported by the correlation and Bland-Altman~\cite{bland_1986} plots shown in Figs.~\ref{fig:correlation}~and~\ref{fig:bland_altman}, respectively.

Fig.~\ref{fig:correlation} shows high correlation between the radar and the Vicon estimates for the stride, step, and stance times, stride and step velocity, stride and step length, and foot maximum velocity. In fact, the correlation coefficients for the stride and step length are the highest reported in the state-of-the-art~\cite{hadjipanayi_2024}. This is beneficial for clinical implementations, which need accurate results. 

The radar network is also robust for home implementations, where analyzing the gait evolution with time is needed. The Bland-Altman plots of Fig.~\ref{fig:bland_altman} show that the estimates are almost unbiased, guaranteeing long-term accuracy.  

Nonetheless, the validation presented in this work is limited to radial gait directions. When gait is non-radial, distinguishing between different body parts becomes more challenging. Therefore, radar nodes should be strategically placed in areas that promote radial movement, such as hallways. If deployed in an unconstrained space, like a living room, the gait analysis would be limited to those gait cycles almost-radial to the radar.

Placing feet nodes at both ends of the corridor may not always be possible. This can be solved by increasing the transmitted power of the radars at one end, as in~\cite{gurbuz_2024, liu_2022, abedi_2022, hadjipanayi_2024}. In those scenarios where it is only possible to implement a single node, the results presented in Table~\ref{tab:mean_error_temp} show that it is better to place it at the torso height and implement the algorithm in~\cite{saho_2022}.

\subsection{Algorithm Limitations}

The contributions shown in Table~\ref{tab:comp_validation} are close to the accuracy limit that can be achieved with the algorithms implemented~\cite{seifert_2019, saho_2022}. The main limitation is linked to the fact that it is not possible to completely isolate the Doppler components of each body part, as all body parts interfere in the Doppler-time matrix. Consequently, a small error is committed in each estimate that affects all the spatiotemporal gait parameters. For instance, when estimating the HS events~\cite{seifert_2019}, it is not possible to observe the exact time frame in which the feet velocity becomes zero, as shown in Fig.~\ref{fig:alg_seifert}. 

Similarly, the feet and torso Doppler contributions are filtered to avoid errors caused by the interferences of other body parts~\cite{seifert_2019, saho_2022}, as shown in Fig.~\ref{fig:alg_saho}. Although necessary, these filters introduce errors into the spatiotemporal gait parameters.

Some radar configuration modifications could improve the algorithm performance by decreasing the interferences of other body parts, or increasing the SNR of the Doppler-time matrix. For instance, antennae with narrower beamwidths would achieve both goals by reducing the field-of-view.

Reducing the radar bandwidth could also improve the SNR: the noise introduced during target integration (see Fig.~3) is proportional to the number of combined range bins, which is reduced when using a narrower bandwidth. The range resolution loss would not alter the system accuracy, since all the gait parameters are extracted from the Doppler-time matrix.

\section{Conclusion}\label{sec:conclusion}

This paper addresses many open questions concerning the deployment of radar systems for gait monitoring. This is achieved by designing a new configurable radar network, analyzing its performance under different configurations, and implementing different algorithms to extract the most relevant parameters.

Furthermore, subjects with PD have been considered for the first time in a radar validation, rising up concerns often overlooked when analyzing healthy subjects.

The results obtained show that:
\begin{itemize}
	\item The data extracted from feet nodes are more relevant to extract temporal gait parameters, especially of aged or motorically impaired subjects.
	\item The data extracted from torso nodes are more relevant to extract the spatial gait parameters.
	\item The most accurate results are obtained by combining two feet and one torso node. To achieve this, a technique to combine Doppler-time matrices has been proposed, which consists on selecting the time frames of the feet nodes with the highest SNR.
	\item The algorithms developed by Seifert et al.~\cite{seifert_2019} are the most suitable to extract the gait parameters.
\end{itemize}

The radar network presented in this paper offers a rapid and accurate analysis of gait. This can be beneficial for clinical and domestic implementations, which need to be time and cost effective. Other state-of-art technologies (cameras, wearables, etc.) are costly and require a large amount of time to set them up and process the data. 

Therefore, radar technology is a reliable solution for gait analysis, robust for aged and motorically impaired people. This is useful to anticipate the detection of movement disorders, or prevent risks linked with age.

\section*{Acknowledegment}
\par{\small 
The authors thank the Neurology Service of Hospital General Universitario Gregorio Marañón, Madrid, and the LAMBECOM laboratory of Universidad Rey Juan Carlos for the facilities provided. Furthermore, the authors thank the professionals at LAMBECOM: Dr. Juan Carlos Miangolarra-Page, Dr. Francisco Molina-Rueda, and Dr. Diego Fernández-Vázquez.

The authors thank the study participants for their availability.
}


\begin{thebibliography}{00}
\bibitem{gurbuz_2024} S.~Z.~Gurbuz, et al., ``Overview of Radar-Based Gait Parameter Estimation Techniques for Fall Risk Assessment,'' \emph{IEEE Open J. Eng. Med. Biol.}, vol. 5, pp.~735-749, Jun. 3 2024.

\bibitem{podsiadlo_1991} D.~Podsiadlo and S.~Richardson, ``The timed "Up \& Go": a test of basic functional mobility for frail elderly persons.'' \emph{J. Amer. Geriatrics Soc.} vol. 39, no. 2, pp. 142-148, 1991.

\bibitem{liu_2022} Y.~Liu, et al., ``Monitoring gait at home with radio waves in Parkinson's disease: A marker of severity, progression, and medication response,'' \emph{Sci. Translational Med.}, vol.~14, p. eadc9669, Sep 2022.

\bibitem{delrobaei_2018} M.~Delrobaei, et al., ``Towards remote monitoring of Parkinson’s disease tremor using wearable motion capture systems'' \emph{J. Neurological Sci.}, vol. 384, pp. 38-45, 2018.

\bibitem{perumal_2016} S.~V.~Perumal and R.~Sankar, ``Gait and tremor assessment for patients with Parkinson’s disease using wearable sensors'' \emph{ICT Exp.}, vol. 2, pp. 168-174, 2016.

\bibitem{zanela_2022} A.~Zanela, et al., ``Using a Video Device and a Deep Learning-Based Pose Estimator to Assess Gait Impairment in Neurodegenerative Related Disorders: A Pilot Study'' \emph{Appl. Sci.}, vol. 12, no. 9, p. 4642, May 2022.

\bibitem{vicon} Vicon Motion Systems Ltd., ``VICON,'' Tech. Rep., 2024.

\bibitem{siva_2024} P.~Siva, et al., ``Automatic Radar-Based Step Length Measurement in the Home for Older Adults Living with Frailty,'' \emph{Sensors}, vol.~24, no.~4, pp.~1056, Feb 2024.

\bibitem{seifert_2019} A.-K. Seifert, M.~G. Amin, and A.~M. Zoubir, ``Toward Unobtrusive In-Home Gait Analysis Based on Radar Micro-Doppler Signatures,'' \emph{IEEE Trans. Biomed. Eng.}, vol.~66, no.~9, pp.~2629-2640, Sep 2019.

\bibitem{kabelac_2019} Z.~Kabelac, et al., ``Passive monitoring at home: A pilot study in parkinson disease,'' \emph{Digit. Biomarkers}, vol.~3, no.~1, pp.~22-30, Apr 2019.


\bibitem{saho_2022} K.~Saho, et al., ``Estimation of gait parameters from trunk movement measured by doppler radar,'' \emph{IEEE J. Electromagn. RF Microw. Med. Biol.}, vol.~6, no.~4, pp.~461-469, 2022.

\bibitem{soubra_2023} R.~Soubra, F.~Mourad-Chehade, and A.~Chkeir, ``Automation of the Timed Up and Go Test Using a Doppler Radar System for Gait and Balance Analysis in Elderly People,'' \emph{J. Healthcare Eng.}, vol.~2023, p. 2016262, 2023.

\bibitem{wang_2014} F.~Wang, et al., "Quantitative Gait Measurement With Pulse-Doppler Radar for Passive In-Home Gait Assessment," \emph{IEEE Trans. Biomed. Eng.}, vol. 61, no. 9, pp. 2434-2443, Sept. 2014.

\bibitem{abedi_2022} H.~Abedi, et al., ``Hallway Gait Monitoring System Using an In-Package Integrated Dielectric Lens Paired with a mm-Wave Radar,'' \emph{Sensors}, vol.~23, no.~1, pp.~71, Dec 2022.

\bibitem{hadjipanayi_2024} C.~Hadjipanayi, et al., ``Remote Gait Analysis Using Ultra-Wideband Radar Technology Based on Joint Range-Doppler-Time Representation,'' \emph{IEEE Trans. Biomed. Eng.}, vol.~71, no.~10, pp.~2854-2865, May 2024.

\bibitem{zeng_2022} X.~Zeng, H.~S.~L.~Báruson, A.~Sundvall ``Walking Step Monitoring with a Millimeter-Wave Radar in Real-Life Environment for Disease and Fall Prevention for the Elderly,'' \emph{Sensors}, vol.~22, no.~24, pp.~9901, Dec. 2022.

\bibitem{wang_2024} L.~Wang, Z.~Ni, B.~Huang ``Extraction and Validation of Biomechanical Gait Parameters with Contactless FMCW Radar,'' \emph{Sensors}, vol.~24, no.~13, pp.~4184, Jun. 2024.

\bibitem{lopezdelgado_2025} I.~E.~López-Delgado, et al., ``Gait symmetry analysis with FMCW MIMO radar,'' \emph{IEEE Trans. Microw. Theor. Techn.}, doi: 10.1109/TMTT.2025.3542183, Feb 2025.

\bibitem{trabassi_2023} D.~Trabassi, et al.``Multiscale entropy algorithms to investigate the degree of complexity and variability of trunk acceleration time series in patients with Parkinson's disease,'' \emph{Gait Posture}, vol.~105, pp.~S49-S50, 2023.

\bibitem{giannakou_2023} E.~Giannakou, et al., ``A Comparative Analysis of Symmetry Indices for Spatiotemporal Gait Features in Early Parkinson's Disease,'' \emph{Neurology Int.}, vol.~15, no.~3, pp.~1129-1139, Sep 2023.

\bibitem{lopez_2023} I.~E. López-Delgado, et al., ``mm-wave wireless radar network for early detection of parkinson's disease by gait analysis,'' in \emph{2023 IEEE Radar Conf. (RadarConf23)}, 2023, pp. 1-6.

\bibitem{stm32wb15cc} ST Microelectronics, Inc., ``STM32WB15CC,'' DS13258 Rev 7. Aug., 2022.

\bibitem{bgt24mtr11} Infineon Technologies AG, ``BGT24MTR11,'' rev.3.1. Mar.25, 2014.

\bibitem{adf4159} Analog Devices, ``ADF4159. Direct Modulation/Fast Waveform Generating, 13 GHz,
Fractional-N Frequency Synthesizer. Datasheet,'' rev.E. 2023.

\bibitem{adib_2015} F.~Adib, Z.~Kabelac, and D.~Katabi, ``Multi-Person localization via RF body reflections,'' in \emph{12th USENIX Symp. Netw. Syst. Des. Implementation (NSDI 15)}, 2015, pp.~279-292.

\bibitem{tivive_2015} F.~H.~C.~Tivive, S.~L.~Phung, and A.~Bouzerdoum, ``Classification of micro-doppler signatures of human motions using log-gabor filters,'' \emph{IET Radar, Sonar Navigation}, vol.~9, no.~9, pp.~1188-1195, 2015.

\bibitem{ash_2018} M.~Ash, M.~Ritchie, and K.~Chetty, ``On the Application of Digital Moving Target Indication Techniques to Short-Range FMCW Radar Data,'' \emph{IEEE Access}, vol.~18, no.~10, pp.~4167-4175, May 2015.

\bibitem{hu_2022} Y.~Hu, and T.~Toda, ``Remote Vital Signs Measurement of Indoor Walking Persons Using mm-Wave FMCW Radar,'' \emph{IEEE Access}, vol.~10, pp.~78219-78230, 2022.

\bibitem{oconnor_2007} C.~M.~O’Connor, et al., ``Automatic detection of gait events using kinematic data,'' \emph{Gait Posture}, vol.~25, pp.~469-474, Mar 2007.

\bibitem{kruskall_1952} W.~H.~Kruskal and W.~A.~Wallis, ``Use of ranks in one-criterion variance analysis,'' \emph{J. Amer. Statist. Assoc.} vol.~47, no.~260, pp.~583–621, Dec 1952.

\bibitem{bland_1986} J.~M.~Bland, and D.~Altman, ``Statistical methods for assessing agreement between two methods of clinical measurement,'' \emph{Lancet}, vol.~327, no.~8476, pp.~307-310, Feb 1986. 
\end{thebibliography}
\end{document}